\documentclass[final]{IEEEtran}
\usepackage[latin9]{inputenc}
\usepackage{color}
\usepackage{amsthm}
\usepackage{amsmath}
\usepackage{amssymb}
\usepackage{graphicx}
\usepackage{setspace}
\usepackage{url}
\PassOptionsToPackage{normalem}{ulem}
\usepackage{ulem}

\usepackage[text={16cm,24cm}]{geometry}
\makeatletter

\providecommand{\tabularnewline}{\\}

\theoremstyle{plain}
\newtheorem{thm}{\protect\theoremname}
\theoremstyle{plain}
\newtheorem{prop}[thm]{\protect\propositionname}
\theoremstyle{plain}
\newtheorem{lem}[thm]{\protect\lemmaname}
\theoremstyle{remark}

\theoremstyle{note}

\renewcommand{\IEEEQED}{\IEEEQEDopen}
\usepackage{cite}
\usepackage{epsfig}
\usepackage{anysize}
\usepackage{algorithm}
\usepackage{algorithmic}
\usepackage{comment}
\usepackage{color}
\marginsize{0.65in}{0.65in}{0.75in}{1in}

\makeatother

\providecommand{\notename}{Note}
\providecommand{\remarkname}{Remark}
\providecommand{\lemmaname}{Lemma}
\providecommand{\propositionname}{Proposition}
\providecommand{\theoremname}{Theorem}

\begin{document}

\title{Cross-Layer Optimization of Fast Video Delivery in Cache-Enabled Relaying Networks}

\author{{Lin Xiang$^*$, Derrick Wing Kwan Ng$^{* \ddag}$, Toufiqul Islam$^\dag$, Robert Schober$^*$, and Vincent W.S. Wong$^\dag$}\\
{$^*$Institute for Digital Communications, Friedrich-Alexander-Universit\"at Erlangen-N\"urnberg, Germany \\
$\dag$Department of Electrical and Computer Engineering, University of British Columbia, Vancouver, BC, Canada}\\
$\ddag$School of Electrical Engineering and Telecommunications, University of New South Wales, Australia \vspace{-.6cm}
}
\maketitle
\pagenumbering{gobble}
\begin{abstract}
This paper investigates the cross-layer optimization of fast video delivery and caching for minimization of the overall video delivery time in a two-hop relaying network. The half-duplex relay nodes are equipped with both a cache and a buffer which facilitate joint scheduling of fetching and delivery to exploit the channel diversity for improving the overall delivery performance. The fast delivery control is formulated as a two-stage functional  non-convex optimization problem. By exploiting the underlying convex and quasi-convex structures, the problem can be solved exactly and efficiently by the developed algorithm. Simulation results show that significant caching and buffering gains can be achieved with the proposed framework, which translates into a reduction of the overall video delivery time. Besides, a trade-off between caching and buffering gains is unveiled.
\end{abstract}

\vspace{-0.35cm}
\section{Introduction}

The surge of video-on-demand (VoD) streaming traffic in cellular networks \cite{Cisco10:Forecast} poses significant challenges for the cellular operators. On the one hand, video streaming imposes stringent requirements on transmission rate and latency \cite{Zhang2014effective}. 
Supporting such resource-intensive applications in cellular networks demands more sophisticated spectrally efficient transmission schemes and more intelligent resource management in the radio access network (RAN) compared to the state of the art. On the other hand, as VoD servers are usually located at the ``Internet edge", a high-capacity backhaul is also required to convey the aggregate VoD data from the Internet to the RAN. However, industry reports suggest that even state-of-the-art 4G LTE networks can suffer from backhaul  capacity bottlenecks during VoD streaming \cite{WSJ13:Verizon}.

Instead of a traditional bottom-up upgrade from RAN to backhaul, cost-effective solutions employing ``caching in the air" (or wireless caching) have been proposed to address the VoD streaming challenges \cite{Wang14:Cache,WCNC12:RANCache,GlobeSIP14:EE}. By pre-storing the most popular files at the base stations (BSs) or access points (APs) of the RAN, wireless caching can achieve significant traffic offloading for the backhaul, delivery capacity enhancement and delay reduction in the RAN \cite{WCNC12:RANCache}, and energy savings in the whole network \cite{GlobeSIP14:EE}. These benefits result from the \emph{content reuse gains} in the video delivery phase, which depend on the correlation between the users' preferences. Besides, the cache placement usually takes place during periods of low cellular traffic to exploit  \emph{traffic diversity gains} and to keep the burden for the system low.

Recently, wireless caching has been integrated into 5G physical layer technologies such as small cells \cite{Caire13IT:FemtoCaching,ICC14:EECaching}, cooperative multiple-input multiple-output (MIMO) \cite{Liu13TSP:CoMP,Liu14TSP:CoMP}, and cross-layer resource allocation schemes \cite{WCNC14:CachingGain:MUdiversity} for advanced cellular video delivery. In \cite{Caire13IT:FemtoCaching}, FemtoCaching was proposed to replace the backhaul in small cell networks, where the optimal file placement was investigated for the minimization of the average download delay 
subject to a cache capacity constraint. In \cite{ICC14:EECaching}, cooperative caching in relay nodes and users was considered for the minimization of the energy consumption. Both  \cite{Caire13IT:FemtoCaching} and \cite{ICC14:EECaching} have shown that caching can effectively relieve the (wireless/wireline) backhaul capacity bottleneck in small cells. Besides, the reduced cell sizes (i.e., smaller BS/cache coverage) lead to additional gains in area spectral efficiency for video delivery. These gains constitute the \emph{macroscopic} caching gains in cache-enabled small cells, which are achievable irrespective of the adopted physical layer.

Meanwhile, caching has been closely combined with the underlying transmission techniques and scheduling protocols to combat wireless channel fading and to alleviate the radio resource scarcity in \cite{Liu13TSP:CoMP,Liu14TSP:CoMP,WCNC14:CachingGain:MUdiversity}. In \cite{Liu13TSP:CoMP}, by caching the same data across different BSs, the authors exploited cooperative MIMO transmission for the minimization of the transmit power under a data rate constraint. Appealingly, the payload sharing overhead of opportunistic cooperative MIMO transmission is reduced by caching, which introduces inexpensive \emph{spatial multiplexing gains}. Because of the low burden for the backhaul, a similar approach was proposed for small cell networks with simple backhaul support \cite{Liu14TSP:CoMP}. In \cite{WCNC14:CachingGain:MUdiversity}, the authors proposed a channel-aware scheduling scheme for transmitting cached data in a one-hop wireless network, which could exploit  \emph{multi-user diversity gains}. These cache-induced benefits in physical layer transmission and resource allocation are referred to as the \emph{microscopic} caching gains. From both the macroscopic and the microscopic perspective, wireless caching is very promising for VoD streaming in future cellular networks.

The aforementioned works \cite{Wang14:Cache,WCNC12:RANCache,GlobeSIP14:EE,Caire13IT:FemtoCaching,ICC14:EECaching,Liu13TSP:CoMP,Liu14TSP:CoMP,WCNC14:CachingGain:MUdiversity} have all considered orthogonal (independent) delivery of cached and uncached video data. Also, the caching policy is designed only for enhancing the delivery of cached data; however, delivering uncached data still suffers from the backhaul capacity limitations. As a result, the user's quality of experience (QoE) may vary significantly when requesting different files. In this paper, the delivery of cached and uncached data is jointly optimized to address the QoE concerns. We investigate cross-layer resource allocation schemes to facilitate overall delivery enhancements in a cache-enabled small cell network with half-duplex relay nodes (RNs).  We assume that each RN is equipped with a cache and a buffer to overcome the backhaul capacity bottleneck and the half-duplex limitation for improving the overall video delivery. 
The roles of the cache and the buffer are summarized below.

The cache is used as a long-term memory to store a certain amount of the video files at the RNs before delivery starts. The cached video files can be fetched directly if they are requested for delivery. In contrast, the buffer is a short-term memory activated during delivery and temporally stores the data packets fetched from both the video server (referred to as ``wireless fetching") and the cache before delivery to the requesting users.

For the delivery of uncached data, the buffer applies buffer-aided relaying (BaR) \cite{Nikola13JSAC:BaR,Nikola14MCOM} to mitigate the rate loss due to the half-duplex operation of the relay. Specifically, the wireless fetching and delivery links are adaptively scheduled in each time slot based on their instantaneous channel state information (CSI). The resulting channel diversity gains can effectively improve the delivery of uncached data. On the other hand, if the requested files are (partially) cached, the cache and the buffer facilitate joint wireless fetching, cache fetching, and delivery control to improve the overall delivery of cached and uncached data. Particularly, the probability of an empty buffer for BaR is decreased as the cached data can feed the buffer with a negligible delay. Thus, the cached data lead to improved diversity gains for delivering the uncached data because of the increased flexibilities in link scheduling.  

Based on the above discussion, what portion of each file is cached should be carefully optimized to best exploit the cache and the buffer for maximization of overall delivery performance. Herein, considering all users' QoE requirements, the minimum delivery time is adopted as the system design objective for enabling \emph{fast} delivery. Our contributions are summarized as follows: 

{\scriptsize$\bullet$} We investigate cross-layer optimization of fast video delivery in a two-hop relay network where both a cache and a buffer are utilized for minimization of the overall delivery time. 

{\scriptsize$\bullet$} We formulate a two-stage optimization problem. In the first stage, the cache status is optimized based on historical profiles of user requests and CSI. In the second stage, cross-layer fast delivery control is performed to minimize the delivery time of the actual requesting users for a given cache status. 

{\scriptsize$\bullet$} The optimization problem is functional (i.e., the feasible set dynamically varies with the values of the optimization variables \cite{Bertsekas1995Dynamic}) and non-convex. By decomposition and transformation, we discover the hidden convexity and quasi-convexity of the underlying subproblems, which allows us to exactly and efficiently solve the problem.

{\scriptsize$\bullet$} Simulation results show that caching and buffering can effectively improve the performance of cellular video delivery by more efficiently utilizing the radio resources. The trade-off between caching and buffering gains is also unveiled.

%The remainder of this paper is organized as follows: We present the system model in Section II. The cross-layer optimization problem is formulated and solved in Section III. In Section IV, simulation results are provided and finally, Section V concludes the paper.   

\vspace{-0.3cm}
\section{System Model}
\label{sec:systmod}
In this section, the system model of the considered cache-enabled two-hop relaying network is presented. 
\vspace{-0.3cm}
\subsection{Cache-Enabled Relay System}
\begin{figure}
\centering\includegraphics[scale=0.54]{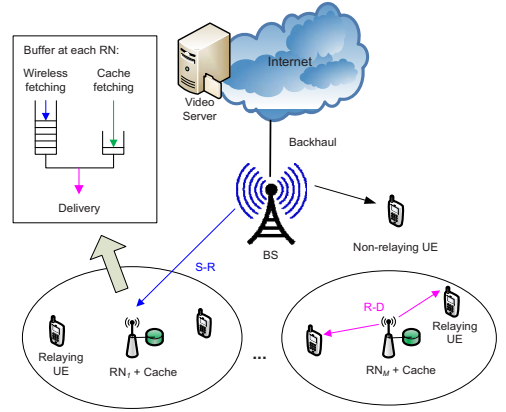}\protect\caption{\label{fig:Coordination-of-Macro}
System model: Caching and buffering at the relay nodes.}
\vspace{-0.5cm}
\end{figure}

We consider a time-slotted video delivery system with time index $t$, where $t\in \mathbb{N}$ is a non-negative integer taken from the set of natural numbers $\mathbb{N}$. The duration of each time slot is $\Delta$. As shown in Figure~\ref{fig:Coordination-of-Macro}, the system consists of a BS overlaid with $M$ half-duplex RNs, which are indexed by $m\in\mathcal{M}=\left\{ 1,\ldots,M\right\} $. Each RN $m$ serves $K^{(m)}$ stationary user equipments (UEs), which belong to set $\mathcal{K}^{(m)}$. We assume that direct communication between the relaying UEs and the BS is not possible due to heavy blockage. The BS fetches video files from the video server on the Internet via the capacity-limited backhaul and delivers them to the (relaying) UEs via the RNs. There are $N$ video files, which have size  $V_{n}$ bits, $n\in\mathcal{N}=\{1,\ldots,N\}$, available on the video server. 

The RNs perform decode-and-forward (DF) relaying and connect the BS and the UEs only via wireless links for deployment convenience. Due to the half-duplex constraint, the RNs alternate between transmitting and receiving. 
Different from conventional systems, each RN is equipped with a cache and a buffer to overcome the backhaul bottleneck and the half-duplex limitation during video delivery. By caching, video files are pre-stored into a secondary memory (e.g. hard disks) before delivery starts and are kept in an indexable file format. In contrast, by buffering, data packets fetched from the cache or video server are input into the primary memory (e.g. Random Access Memory (RAM) and Dynamic RAM) and output from the primary memory during delivery to the UEs in a first-in first-out (FIFO) manner.

The system operation has two stages: cache placement and delivery. During placement, video files are proactively stored into the cache. We assume that the users' preferences vary slowly. Then, cache placement can be updated at a low frequency, e.g. during the off-peak network traffic periods in the early mornings. The caching decision process and the cache placement are assumed to be completed before delivery starts at time $t=0$.

Assuming that each chunk of the video file is encoded via rateless maximum distance separable codes \cite{Caire13IT:FemtoCaching}, we can cache $c_{n}^{(m)}\in [0,1]$ portion of file $n$ into RN $m$ at $t=0$. Denote the cache allocation vector normalized with respect to the file sizes by $\mathbf{c}^{(m)}=[c_{1}^{(m)},\ldots,c_{N}^{(m)}]$ at RN $m$.  The cache placement cannot exceed the cache capacity $C_{\max}^{(m)}$ at RN $m$, i.e.,
\begin{equation}
\begin{aligned}\textrm{C1: } & \sum\nolimits_{n=1}^{N}c_{n}^{(m)}V_{n}\le C_{\max}^{(m)},\;\forall m \in \mathcal{M}.\end{aligned}
\label{eq:cache-memory}
\end{equation}

In the delivery stage, each user may request one video file at $t=0$. Let $\mathcal{K}_{n}^{(m)}$ be the set of users connected to RN $m$ requesting file $n$. We use the tuple $\boldsymbol{\rho}\equiv(m,k,n)$ to represent a request for file $n$ by user $k$ belonging to RN $m$. The set of all requests of the UEs is denoted by $\mathcal{G} \subseteq \bigcup\nolimits_{m=1}^{M} \{\mathcal{K}_{n}^{(m)} \times \mathcal{N} \}$, where $\mathcal{X} \times \mathcal{Y}$ is the Cartesian product of sets $\mathcal{X}$ and $\mathcal{Y}$, i.e., we have  $\boldsymbol{\rho}\in\mathcal{G}$. When a file is requested, the RN searches the file indices in its cache. If RN $m$ has cached file $n$, i.e., $c_{n}^{(m)} > 0$, $c_{n}^{(m)}V_{n}$ bits of file $n$ can be fetched directly from the cache, while the rest has to be fetched from the video server, which is referred to as wireless fetching, for correct decoding of the video chunks \cite{Caire13IT:FemtoCaching}. Otherwise, the entire file has to be obtained by wireless fetching.  

\vspace{-.3cm}
\subsection{Joint File Fetching and Delivery }
\label{subsec2-2}
\vspace{-.1cm}
In the considered two-hop network, the wireless BS-to-RN$_{m}, m \in \mathcal{M}$, fetching links and the RN$_{m}$-to-UE$_{k}$, $k\in\mathcal{K}_{n}^{(m)}$, delivery links constitute the source-to-relay ($\mathcal{S}-\mathcal{R}$) and the relay-to-destination ($\mathcal{R}-\mathcal{D}$) links, respectively. If the requested files are uncached, the buffers equipped at the RNs facilitate BaR \cite{Nikola13JSAC:BaR,Nikola14MCOM} to overcome the half-duplex limitation, where the $\mathcal{S}-\mathcal{R}$ and $\mathcal{R}-\mathcal{D}$ links are scheduled opportunistically in each time slot to maximize the data rate. If the requested files are (partially) cached, joint (wireless and cache) fetching and delivery is possible to cooperatively control the cache and the buffer for improving the overall delivery. 

To investigate joint (wireless and cache) fetching and delivery in the considered multi-RN multi-user system, orthogonal frequency division multiplex access (OFDMA) is applied at the physical layer, which can effectively combat frequency-selective fading. We  consider cross-layer resource allocation, where joint adaptive subcarrier 
(SC) assignment, link scheduling, and power allocation is performed at the physical/link layer for fast video delivery with a minimum delivery rate guarantee. 

\subsubsection{Joint SC Assignment, Link Scheduling, and Power Allocation}
The RNs are allowed to switch between transmitting and receiving within each time slot by time-division duplexing. Hence, we assume that the $\mathcal{S}-\mathcal{R}$ and  the $\mathcal{R}-\mathcal{D}$ links are activated for $\eta_{S,t}^{(m)}\in[0,1]$ and $\eta_{R,t}^{(m)}\in[0,1]$ fractions of time slot $t$, respectively, where $\eta_{S,t}^{(m)}+\eta_{R,t}^{(m)} = 1,\;\forall t$.

On the other hand, the available frequency spectrum consists of $F$ orthogonal SCs, each of bandwidth $W$. The SCs are assigned to each $\mathcal{S}-\mathcal{R}-\mathcal{D}$ path in each time slot. Let binary variable $\mu_{\boldsymbol{\rho},f,t}\in\left\{ 0,1\right\} $ indicate the assignment of SC $f \in \mathcal{F} = \left\{ 1,\ldots,F\right\} $ to the $\mathcal{S}-\mathcal{R}-\mathcal{D}$ path for serving request $\boldsymbol{\rho}\in \mathcal{G}$ at time $t$. Due to time switching, each SC is split into two subchannels at time $t$, which are indexed by $(f,t,i)$, $i \in \{S, R\}$, and used on the $\mathcal{S}-\mathcal{R}$ and $\mathcal{R}-\mathcal{D}$ links, respectively.  Let $\mu_{\boldsymbol{\rho},f,t}^{i}$ be the effective allocation of subchannel $(f,t,i)$ in serving request $\boldsymbol{\rho} $. We have 
\begin{equation}
\begin{aligned} \textrm{C2: } &  \mu_{\boldsymbol{\rho},f,t}=\mu_{\boldsymbol{\rho},f,t}^{S}+\mu_{\boldsymbol{\rho},f,t}^{R},\quad\forall\boldsymbol{\rho},f,t,\\
\textrm{C3: } &0\le\mu_{\boldsymbol{\rho},f,t}^{i}\le\eta_{i,t}^{(m)}, \quad \forall\boldsymbol{\rho},f,t,i\in\left\{ S,R\right\},\\
\textrm{C4: } & \mu_{\boldsymbol{\rho},f,t}\in\left\{ 0,1\right\},  \quad\forall\boldsymbol{\rho},f,t,\\
\textrm{C5: } & \eta_{S,t}^{(m)} + \eta_{R,t}^{(m)} = 1, \quad \forall t,m,\\
\end{aligned}
\label{eq:tfu}
\end{equation}
where C3 coordinates the switching time in one RN. As a result, the $\mathcal{S}-\mathcal{R}$ link is active if $\eta_{S,t}^{(m)}>0$ and $\mu_{\boldsymbol{\rho},f,t}^{S}>0$; similarly, the $\mathcal{R}-\mathcal{D}$ link is active if $\eta_{R,t}^{(m)}>0$ and $\mu_{\boldsymbol{\rho},f,t}^{R}>0$.

Consider flat fading on each SC where the duration of a time slot is less than the channel coherence time. Let $h_{\boldsymbol{\rho},f,t}^{S}$ and $h_{\boldsymbol{\rho},f,t}^{R}$ be the instantaneous channel states  of the $\mathcal{S}-\mathcal{R}$ and the $\mathcal{R}-\mathcal{D}$ links on SC $f\in\mathcal{F}$ in time slot $t$, respectively. Moreover, $p_{f,t}^{S}$ and $p_{f,t}^{R}$ are the transmit powers allocated to the corresponding subchannels. Then, the following constraints are considered for SC assignment and power allocation:
\begin{equation}
\begin{aligned}\textrm{C6: } & \sum\nolimits_{f}\mu_{\boldsymbol{\rho},f,t}\le1,\quad\forall\boldsymbol{\rho} \in \mathcal{G},\\
\textrm{C7: } & \sum\nolimits_{\boldsymbol{\rho}}\mu_{\boldsymbol{\rho},f,t}\le1,\quad\forall f \in \mathcal{F},\\
\textrm{C8: } & \sum\nolimits_{\boldsymbol{\rho},f}  \mu_{\boldsymbol{\rho},f,t}^{i}p_{f,t}^{i}  \le  P_{i}, \;\; \forall t, i \in \{S, R\},\\
\end{aligned}
\label{eq:sc-pwr}
\end{equation}
where C6 and C7 guarantee that each SC is assigned only once in a time slot and to at most one $\mathcal{S}-\mathcal{R}-\mathcal{D}$ path; and C8 constrains the instantaneous transmit power of the $\mathcal{S}-\mathcal{R}$ and $\mathcal{R}-\mathcal{D}$ links to $P_{S}$ and $P_{R}$, respectively.

\subsubsection{Cache/Buffer Management at the RNs}

As illustrated in Figure~\ref{fig:Coordination-of-Macro}, we assume the buffer uses two FIFO queues for wireless fetching and cache  fetching, respectively. For joint fetching and delivery, the two queues are cooperatively controlled in accordance with the resource allocation policy to exploit the channel diversity.

Let $B_{\boldsymbol{\rho},t}^{C}$ and $B_{\boldsymbol{\rho},t}^S$ be the accumulated data that has to be fetched from the cache and the video server, respectively, and $B_{\boldsymbol{\rho},t}^R$ be the accumulated delivered data in serving request $\boldsymbol{\rho}$. For given $h_{\boldsymbol{\rho},f, t}^{i}$ and $p_{f, t}^{i}$, the capacity of subchannel $(f,t,i)$ is $ \log_{2} (1+p_{f, t}^{i}h_{\boldsymbol{\rho},f, t}^{i}/(N_0 W) )$ bps/Hz, where $N_0$ is the noise power spectral density and $i \in \{S, R\}$. Thus, for a given resource allocation decision, the queues evolve as 
\vspace{-.2cm}
\begin{equation*}
\begin{aligned}
B_{\boldsymbol{\rho},t}^{i} & =W\Delta  \sum_{\tau=1}^{t}\sum_{f \in \mathcal{F}} \mu_{\boldsymbol{\rho},f, \tau}^{i}\log_{2}\Big(1+\frac{p_{f, \tau}^{i}h_{\boldsymbol{\rho},f, \tau}^{i}}{N_0 W} \Big), \, i \in \{S, R\},\\
B_{\boldsymbol{\rho},t}^{C} & =\Delta \sum\nolimits_{\tau=1}^{t}b_{\boldsymbol{\rho},\tau}^{C},\;\forall t,\,\boldsymbol{\rho},\\
\end{aligned}
\label{qevolve}
\end{equation*}
where $b_{\boldsymbol{\rho},\tau}^{C}\ge0$ is the instantaneous rate of cache fetching at time $\tau$. Meanwhile, the queue evolution needs to satisfy the following boundary constraints: 
\begin{equation}
\begin{aligned}\textrm{C9: } & B_{\boldsymbol{\rho},t}^R\le \min \left\{ B_{\boldsymbol{\rho},t}^S +B_{\boldsymbol{\rho},t}^{C}, \;V_n\right\},\;\forall\boldsymbol{\rho},t,\\
\textrm{C10: } & B_{\boldsymbol{\rho},t}^{C}\le c_{n}^{(m)}V_{n},\;\forall\boldsymbol{\rho},t,\\
\textrm{C11: } & \sum\nolimits_{\boldsymbol{\rho}}\left(B_{\boldsymbol{\rho},t}^S + B_{\boldsymbol{\rho},t-1}^{C}- B_{\boldsymbol{\rho},t-1}^R \right)\le B_{\max}^{(m)},\;\forall m,t, \\
\textrm{C12: } & \sum\nolimits_{m, \boldsymbol{\rho}} \left( B_{\boldsymbol{\rho},t}^S - B_{\boldsymbol{\rho},t-1}^S \right) \le   \Gamma_{t} \Delta ,\;\forall  t,
\end{aligned}
\end{equation}
where C9 and C10 guarantee data causality in the buffer/cache, i.e., the amount of delivered data cannot exceed  the aggregate cached/buffered data nor the file size; C11 constrains the total queue length at the RNs by the buffer capacity $B_{\max}^{(m)}$; and C12 limits the sum rate of wireless fetching to be within the backhaul capacity $\Gamma_t$ at each time $t$, i.e., the backhaul is shared by all $\mathcal{S}-\mathcal{R}$ links for video fetching from the video server.

\subsubsection{Ahead-of-Time Video Delivery}
Our scheme is applicable for ahead-of-time video delivery at the application layer, where UEs play the video back after delivery is completed, e.g.  in download-and-play video services. In this case, UEs have a large enough memory to store the downloaded data and assemble them back into the original data sequence before decoding. As a quality of service (QoS) constraint, the cross-layer controller maintains a minimum delivery rate  $\nu_{\min}$ at the UEs, 
\begin{equation}
\textrm{C13: } B_{\boldsymbol{\rho},t}^R \ge \nu_{\min}\cdot \left\{ \min \left\{t-\epsilon_{\boldsymbol{\rho}}, 0\right\} \right\}, \forall\boldsymbol{\rho},t,
\end{equation}
where $\epsilon_{\boldsymbol{\rho}}$ is the initial delivery delay and is known to the controller through feedback. When $t > \epsilon_{\boldsymbol{\rho}}$, C13 reduces to $B_{\boldsymbol{\rho},t}^R / (t-\epsilon_{\boldsymbol{\rho}}) \ge \nu_{\min}$, i.e., a minimum time-averaged delivery rate is maintained at each time.

Although the above model focuses on relaying UEs, it is general enough to also include non-relaying UEs which communicate with the BS directly, cf. Figure~\ref{fig:Coordination-of-Macro}. If user $k'$ is served by the BS, we can configure a ``virtual'' RN $m'$  to serve user $k'$ where $h_{\boldsymbol{\rho}',t}^{S}=h_{\boldsymbol{\rho}',t}^{R}$ and $B_{\boldsymbol{\rho}',t} ^S = B_{\boldsymbol{\rho}',t} ^R $. Based on this formulation, the one-hop communication between non-relaying UEs and the BS can be included as a special case of the considered two-hop communication model. 

\vspace{-0.3cm}
\section{Cross-Layer Caching and Delivery Control}

In this section, the cross-layer caching and delivery control is formulated as a two-stage optimization problem. In the first stage, the cache controller optimizes the cache status based on historical profiles of user requests and CSIs. Then, when the actual user requests arrive, the cross-layer delivery control is applied in the second stage, where all activities concerning the delivery in the different layers are aligned to minimize the overall delivery time. In the following, however, our discussion starts from the simpler second-stage control before tackling the first-stage problem. In this paper, we focus on offline optimization, which assumes full knowledge of the CSIs of all links over time. The solution of the offline problem provides a performance upper bound for the minimum video delivery time and can serve as a starting point for the design of online delivery schemes. 

\vspace{-0.3cm}
\subsection{Second Stage Fast Delivery Control}

For a given cache status, the \emph{delivery time} is defined as the number of time slots needed to complete the file delivery for all users. Denote the delivery time by $T_{\textrm{a}} $, where $T_{\textrm{a}} \in \mathbb{N}$. 
Let  
$\mathbf{d}_{t} (T_{\textrm{a}})=[\mu_{\boldsymbol{\rho},f,t},\mu_{\boldsymbol{\rho},f,t}^{S},\mu_{\boldsymbol{\rho},f,t}^{R},\eta_{S,t}^{(m)},\eta_{R,t}^{(m)},p_{f,t}^{S},p_{f,t}^{S}, b_{\boldsymbol{\rho},t}^{C}]$ be the resource allocation vector belonging to the feasible set $\mathcal{D}$,
\[
\mathcal{D} (T_{\textrm{a}}) \triangleq \left\{\mathbf{d}_{t} \mid \mathbf{d}_{t} \succeq \mathbf{0},  \textrm{  C2--C13} \right\}, 
\]
where {{$\mathbf{a} \succeq \mathbf{0}$ indicates that each element of $\mathbf{a}$ is nonnegative}}.  Both $\mathbf{d}_{t} $ and  $\mathcal{D}$ are dependent on the delivery time $T_{\textrm{a}}$.

After the users' requests become known, the cross-layer scheduler computes the optimal delivery vector $\mathbf{d}_{t}$ that minimizes $T_{\textrm{a}}$ subject to the constraints in the different layers. The offline delivery time minimization problem is defined as,
\begin{equation}
\begin{aligned}\textrm{(P1a)}\quad \min_{T_{\textrm{a}}\in \mathbb{N},  \; \mathbf{d}_{t} \in \mathcal{D}}\quad & T_{\textrm{a}} \\
\textrm{s.t.}\;\,\qquad  & \textrm{C14: } B^R_{\boldsymbol{\rho},T_{\textrm{a}}} = V_{n},\;\forall\boldsymbol{\rho},
\end{aligned}
\end{equation}
where constraint C14 indicates delivery completion at $T_{\textrm{a}} $.

Problem (P1a) is a mixed-combinatorial and non-convex optimization problem, due to the integer variables $\mu_{\boldsymbol{\rho},f,t}$ and $T_{\textrm{a}}$ and the non-convex constraints C8, which contains bilinear terms, and C9, C10, C11, C12, which involve differences of convex functions\footnote{The term on the right-hand side of (C9) is an affine function and also a convex function. Due to the differences of convex functions, the time relaxation and problem transformation in \cite{KwanTCOM12} is not applicable here. Besides, although C14 is also non-convex, 
an equivalent convex constraint can be easily obtained as $B^R_{\boldsymbol{\rho},T_{\textrm{a}}} \ge V_{n}$.}. This type of problem is generally intractable \cite{Horst2000Global}. Meanwhile, the feasible set and the delivery decision variables in (P1a) dynamically vary with $T_{\textrm{a}}$, which is an optimization variable rather than a fixed quantity. This endows (P1a) with the functional attribute  \cite{Bertsekas1995Dynamic} and further complicates the problem. However, we will overcome both difficulties through suitable decomposition and transformation techniques  below.

First, rather than solving (P1a) directly, we decompose it into two subproblems. Let us investigate a ``dual'' problem of (P1a)---the effective throughput maximization in a given delivery time $[0,T_{\textrm{b}}]$, as defined by 
\begin{equation}
\textrm{(P1b)}\quad \begin{aligned}\phi ({T_{\textrm{b}}})=  \max_{ \mathbf{d}_{t} \in \mathcal{D} }\quad & \sum\nolimits_{\boldsymbol{\rho}}  B^R_{\boldsymbol{\rho},T_{\textrm{b}}} ,
\end{aligned}
\label{p1b}
\end{equation}
where the optimal value is denoted by function $\phi (\cdot)$ of $T_{\textrm{b}}$. The relation between (P1a) and (P1b) is stated in the following proposition. 
\begin{prop}
\label{prop:1}
\normalfont
(P1a) and (P1b) are dual in the sense that 
\begin{equation}
T_{\textrm{a}}^*=T_{\textrm{b}}^{*} \triangleq \min_{T_{\textrm{b}}\in \mathbb{N}} \;   T_{\textrm{b}} + \mathbb{I}_{\mathcal{A} } ({T_{\textrm{b}}}) ,
 \label{outerp1}
\end{equation}
where $T_{\textrm{a}}^*$ is the optimal completion time of (P1a) and $\mathbb{I} _{\mathcal{ A}} (x)$  is the indicator function defined on set $\mathcal{A} \triangleq \big\{ x \mid  \phi (x) = \sum\nolimits_{\boldsymbol{\rho}}V_{n} \big\} $, 
\begin{equation*}
\mathbb{I} _{\mathcal{ A}} (x) \triangleq  \Big\{ {\begin{array}{cl}
0, & \quad \textrm{if $x \in \mathcal{A}$ }, \\
\infty, & \quad \textrm{otherwise}.\end{array}}
\end{equation*}
Besides, over set $\mathcal{A}' \triangleq \big\{  x \mid \sum\nolimits_{\boldsymbol{\rho}}  B^R_{\boldsymbol{\rho},x}  = \sum\nolimits_{\boldsymbol{\rho}}V_{n} \big\} $, we have
\begin{equation}
T_{\textrm{a}}^*= T_{\textrm{c}}^*  \triangleq \min_{T_{\textrm{c}}\in \mathbb{N}, \, \mathbf{d}_{t} \in \mathcal{D}} \;  T_{\textrm{c} }+  \mathbb{I}_{\mathcal{A}'} ({T_{\textrm{c}}}).
\label{eq2-prop1}
\end{equation}
 \end{prop}
\begin{IEEEproof}
We first prove (\ref{eq2-prop1}). It is easy to verify that the completion condition C14 is equivalent to 
$\mathbb{I} _{\mathcal{ A}'} (T) = 0$ or $\mathcal{ A}' \neq \emptyset$, since $\mathbb{I} _{\mathcal{ A}'} (T) = \sum\nolimits_{\boldsymbol{\rho}} {\mathbb{I} _{\mathcal{ A}_{\boldsymbol{\rho}}} (T)}$, 
where 
$\mathcal{ A}_{\boldsymbol{\rho}} (T) \triangleq \{T \mid T \textrm{ satisfying C14 for } {\boldsymbol{\rho}}\}$. 
When the completion condition is satisfied, the objective functions and the feasible sets of (P1a) and (\ref{eq2-prop1}) become the same. Thus, (\ref{eq2-prop1})  holds. 

Now, we can prove  (\ref{outerp1}) based on (\ref{eq2-prop1}). From (P1b), ${\mathcal{ A}} (T) \subseteq   {\mathcal{ A}'} (T) $ holds for any $T \in \mathbb{N}$. Then, we have $T_{\textrm{b}}^*\ge T_{\textrm{c}}^* $ since the feasible set of  (\ref{eq2-prop1}) contains that of (\ref{outerp1}). However, if $T_{\textrm{c}}^* \in {\mathcal{ A}'} \neq \emptyset$ and $\mathbf{d}_t^*$ solves (\ref{eq2-prop1}),  then $\mathbf{d}_t^*$ is also feasible for (P1b) with $T_{\textrm{b}} = T_{\textrm{c}}^* \in {\mathcal{ A}'}$. Thus, $T_{\textrm{c}}^* \in {\mathcal{ A}}$. We have $T_{\textrm{b}}^* \le T_{\textrm{b}} = T_{\textrm{c}}^*$ since  $T_{\textrm{b}}^*$ is the optimal solution to (\ref{outerp1}). Therefore, $T_{\textrm{b}}^* = T_{\textrm{c}}^*$ holds, which completes the proof.
\end{IEEEproof}

Based on (\ref{outerp1}), (P1a) can be decomposed into two subproblems: the inner problem (P1b), which maximizes the effective throughput for a given delivery time, and the outer problem (\ref{outerp1}), which seeks the optimal delivery time. Similar to (P1a), the inner subproblem (P1b) is a mixed-combinatorial and non-convex optimization problem. In Section~\ref{sec:iii-c}, we show that (P1b) can be further transformed into an equivalent convex problem which can be efficiently solved. As for the outer subproblem, it can be solved via a simple one-dimensional search. 
Particularly, an efficient bisection method is applicable due to the underlying quasi-convexity of the outer subproblem and the solution is globally optimal for (P1a) under mild conditions \cite{Boyd2004Convex}. The solution of both subproblems will be provided in Section~\ref{sec:iii-c}.

\vspace{-0.5cm}

\subsection{First Stage Cache Control} 
Caching popular files has been considered in the literature mainly for improving the delivery capacity and reducing energy consumption \cite{WCNC12:RANCache,GlobeSIP14:EE}. However, this can lead to great variations in the users' QoEs across different requests. Also, the delivery of uncached files still suffers from capacity limitations. For applications such as video delivery, failure to satisfy the QoE requirement can lead to the loss of customers, which is undesirable for video service providers. Therefore, it is desirable to leverage the caching gains for providing ubiquitous delivery service with fairness guarantees across individual QoEs.

The optimal cache status is determined based on historical profiles of user requests and CSIs, which are referred to as ``scenario" data.  We assume that each user requests only one file at a time and $\Omega$ sets of scenario data can be obtained from the system records. The  delivery decision for scenario $\omega \in \{1, \ldots, \Omega\}$ is denoted by $\mathbf{d}_{t,\omega}$ with the corresponding feasible set $\mathcal{D}_{\omega} (T_{\omega})$. The caching decisions $\mathbf{c}^{(m)} $ are, however, scenario independent and have the feasible set 
 \[ \mathcal{C}^{(m)}=\{\mathbf{c}^{(m)} \mid \mathbf{c}^{(m)} \in [0,1]^N, \textrm{ C1}\}. \]
For providing fairness in QoE, the caching control aims to minimize the worst-case delivery time over the $\Omega$ preselected scenarios, 
\begin{equation}
\begin{aligned}\textrm{(P2a)} \quad  \min_{ T_{\omega} \in \mathbb{N}, \; \mathbf{c}^{(m)} \in \mathcal{C}^{(m)} }   \quad & \max_{\omega=1,\ldots,\Omega} \quad T_{\omega} \\
\textrm{s.t.}\,\quad\qquad\; & \textrm{C15: } B^R_{\boldsymbol{\rho},T_{\omega},\omega} = V_{n},\;\forall\boldsymbol{\rho}, \omega, \\
& \mathbf{d}_{t,\omega} \in \mathcal{D}_{\omega}, \; \forall \omega, 
\end{aligned}
\end{equation}
where C15 indicates delivery completion for each scenario.

Similar to (P1a), we can decompose (P2a) into two subproblems as summarized in the following proposition. 
\begin{prop}
\label{prop:3}
\normalfont
Let $T^*$ be the optimal delivery time of (P2a) and $\varphi (T)$ be the optimal throughput within $[0,T]$ for any given $T \in \mathbb{N}$,
\begin{equation}
\begin{aligned} \textrm{(P2b)} \quad   \varphi (T) = \max_{\mathbf{c}^{(m)} \in \mathcal{C}^{(m)}  }  \quad & \sum_{\boldsymbol{\rho}, \; \omega}  B_{\boldsymbol{\rho}, T, \omega}^R \\
\textrm{s.t.} \qquad & \mathbf{d}_{t,\omega} \in \mathcal{D}_{\omega}, \; \forall \omega. \\ 
\end{aligned}
\end{equation}
Then, over set $\mathcal{B} \triangleq \big\{ x \mid  \varphi (x) = \sum\nolimits_{\boldsymbol{\rho}, \omega}V_{n} \big\} $, we have 
\begin{equation}
T^*= \min_{T \in \mathbb{N}} \;   T  + \mathbb{I}_{\mathcal{B} } (T) ;
\label{outerp2}
\end{equation}
furthermore, over set  $\mathcal{B}' \triangleq \big\{  x \mid \sum\nolimits_{\boldsymbol{\rho}, \omega}  B^R_{\boldsymbol{\rho},x, \omega}  = \sum\nolimits_{\boldsymbol{\rho}, \omega}V_{n} \big\} $, 
\begin{equation}
T ^* = \min_{T \in \mathbb{N}} \; T  + \mathbb{I}_{\mathcal{B}'} (T) .
\label{outerp2-2}
\end{equation}
\end{prop}
\begin{IEEEproof}
Let $\max_{\omega} T_{\omega} \le T$ hold for any $\omega \in \Omega$. By applying the epigraph transformation to (P2a), we obtain (P2b).  Then, (\ref{outerp2}) and (\ref{outerp2-2}) can be proved in a similar manner as Proposition~\ref{prop:1} and the details are omitted here. 
\end{IEEEproof}

\vspace{-0.3cm}
\subsection{Solution of (P1a) and (P2a)}
\label{sec:iii-c}

\subsubsection{Solving Inner Subproblems (P1b) and (P2b)}
For both subproblems,  we apply the constraint relaxation method to deal with the combinatorial variables $\mu_{\boldsymbol{\rho},f,t}$ and variable transformations to address the non-convex constraints C8, C9, C10, C11, and C12. The solution is summarized in two steps.

\emph{Step 1: Binary relaxation for the SC assignment variables: } We relax the binary SC assignment variables into the real domain, i.e.,  $\mu_{\boldsymbol{\rho},f,t}\in[0,1]$. Such a relaxation simplifies the computational process of combinatorial problems. Moreover, the relaxation is tight (optimal) when the number of SCs is large, i.e., $F\to\infty$, as has been proved in \cite{Yu2006Dual}. 

\emph{Step 2: Equivalent convex problem:} We introduce two new variables $\tilde{b}_{\boldsymbol{\rho},f,t}^S$ and $\tilde{b}_{\boldsymbol{\rho},f,t}^R$ which denote the effective fetching and delivery rates on subchannel $(f,t,i)$, respectively, i.e., 
\begin{equation}
\begin{aligned}
\tilde{b}_{\boldsymbol{\rho},f,t}^i = W \Delta  \mu_{\boldsymbol{\rho},f,t}^{i} \log_{2} \Big(1 + \frac{p_{f,t}^{i} h_{\boldsymbol{\rho},f,t}^{i}}{N_0 W} \Big) \ge 0, \; i \in \{S,R\}. \\
\end{aligned}
\label{transform1}
\end{equation}
We have $\tilde{b}_{\boldsymbol{\rho},f,t}^i \to 0$ when ${\mu}_{\boldsymbol{\rho},f,t}^i  \to 0$. Eliminating $p_{f,t}^{S}$ and $p_{f,t}^{R}$ based on (\ref{transform1}), C8 becomes, 
\[
\begin{aligned} & \sum_{\boldsymbol{\rho}, f}  \frac{\mu_{\boldsymbol{\rho},f,t}^{i}}{h_{\boldsymbol{\rho},f,t}^{i}}\left[\exp\left(\frac{\tilde{b}_{\boldsymbol{\rho},f,t}^i \ln 2}{\mu_{\boldsymbol{\rho},f,t}^{i} W \Delta}\right)-1\right] \le \frac{P_{i}}{W N_0}, \forall t, i \in \{S,R\}, \\
\end{aligned}
\]
which is a convex set, cf. \emph{Lemma \ref{lem:7}.}
\begin{lem}
\label{lem:7}
\normalfont
The function $g(ax+b,y)=\left(ax+b\right)\cdot\exp\left(\frac{y}{ax+b}\right)$
is jointly convex in $(x,y)\in\left\{ x|ax+b\ge0\right\} \times\mathbb{R}_{+}$, where $\mathbb{R}_{+}$ is the set of non-negative real numbers.
 \end{lem}
\begin{IEEEproof}
When $a=1$, $b=0$, $g(x,y)$ is a perspective function and thus jointly convex in $(x,y) \in \mathbb{R}_{+}^2$. Then, $g(ax+b,y)$  is the composition of $g(x,y)$ and an affine function, which is also jointly convex  \cite{Boyd2004Convex}. 
\end{IEEEproof}
{{Applying the transformation (\ref{transform1}), $B_{\boldsymbol{\rho},t}^S$ and $B^R_{\boldsymbol{\rho},t}$ become
\begin{equation}
\begin{aligned}B_{\boldsymbol{\rho},t}^i & =\sum\nolimits_{\tau=1}^{t} \sum\nolimits_f\tilde{b}_{\boldsymbol{\rho},f,\tau}^i, \; \forall t, i \in \{S, R\},
\end{aligned}
\label{eq:15}
\end{equation}
which are just affine functions of $\tilde{b}_{\boldsymbol{\rho},f,t}^S$ and $\tilde{b}_{\boldsymbol{\rho},f,t}^R$, respectively.}} As a result, the cache/buffer management constraints C9, C10, C11, and C12 become convex. Therefore, based on these relaxation and transformation steps, (P1b) and (P2b) can be reformulated as equivalent convex problems for which strong duality holds and efficient polynomial-time algorithms exist, e.g., interior point methods \cite{Boyd2004Convex}, to obtain the optimal solution.

\subsubsection{Solving Outer Subproblems (\ref{outerp1}) and (\ref{outerp2})}
The outer subproblems, which are shown to be quasi-convex programs in the following, can be solved by the bisection method. For a strict mathematical proof, let us first extend the domain of delivery time $T$ onto $\mathbb{R}_+$ by defining $T= \lceil t_c\rceil$, where $t_c \in \mathbb{R}_+$ and $\left \lceil \cdot \right \rceil$ is the ceiling function. It is easy to verify that $\phi(\cdot)$ and $\varphi(\cdot)$ defined in (P1b) and (P2b) are non-decreasing and quasi-linear in $t_c$ and $T$. Then, $\mathcal{A}$ and $\mathcal{B}$ are convex sets (specifically rays) in $t_c$. By checking the quasi-convexity of the objective function of (\ref{outerp1}) and (\ref{outerp2}) in $T$, the outer subproblems can be shown to be quasi-convex programs (but not necessarily strictly quasi-convex), for which bisection is an efficient method to obtain the optimal solution \cite{Boyd2004Convex}.

The overall solution procedure for (P1a) and (P2a) is summarized in Algorithm~\ref{alg1}. The algorithm includes a doubling search (i.e., the search interval is doubled in each iteration before termination) from line \ref{alg1:line2} to line \ref{alg1:line7} which determines an upper bound on the delivery time (i.e., an initial search range $[l_0, u_0]$) and a bisection search from line  \ref{alg1:line8} to line \ref{alg1:line17} which optimizes the delivery time. Besides, during each iteration of the search,  the inner problems (\ref{outerp1}) and (\ref{outerp2}) need to be solved \cite{Boyd2004Convex}. For an initial step size $T_{\textrm{step}} = 1$ and delivery time bound $[l_0, u_0]$, both the doubling and bisection search terminate after $\left \lceil  \log_2(u_0 - l_0) \right \rceil$ iterations. Although (\ref{outerp1}) and are not strictly quasi-convex, the bisection method will always find the unique global optimal delivery time under mild conditions, e.g. if $l_0$ is small enough such that $\mathbb{I} _{\mathcal{ A}} (l_0) = \mathbb{I} _{\mathcal{B}} (l_0) = \infty$ (in other words, video delivery does not complete at $T = l_0$). Upon obtaining the optimal delivery time, the corresponding caching or delivery decisions are also available. Thus, the solution procedure for (P1a) and (P2a) is complete. Note that the hidden convexity of the inner subproblem guarantees that the search region and subregions are always convex during the bisection search and the overall solution is globally optimal. 

\vspace{.5cm}

\begin{algorithm}[t]
\caption{\textcolor{black}{Search for the Optimal Delivery Time} }
\label{alg1}
\begin{algorithmic}[1]
\STATE \textbf{Initialization}: \textbf{Given} $T$;  $T_{\textrm{step}} \leftarrow 1$, $l \leftarrow 0$; 

\STATE \emph{\%Phase 1: Doubling search for delivery time bound;} \label{alg1:line2}
\REPEAT
\STATE  Solve inner problems  (P1b) and (P2b) in $[0, T]$;

\STATE $l \leftarrow T$, $T \leftarrow T+ T_{\textrm{step}} $, $T_{\textrm{step}} \leftarrow 2 * T_{\textrm{step}}$; \label{alg1:line5}

\UNTIL{$\mathbb{I} _{\mathcal{ A}} (T) = 0$ or $\mathbb{I} _{\mathcal{B}} (T) = 0$.}

\STATE $u \leftarrow T$; \label{alg1:line7}

\STATE \emph{\%Phase 2: Bisection search for optimal delivery time;} \label{alg1:line8}

\REPEAT
\STATE $T \leftarrow  \left \lceil  (l+u)/2 \right \rceil $;\label{alg1:line8}
\STATE  Solve inner problems  (P1b) and (P2b) in $[0, T]$;

\IF{$\mathbb{I} _{\mathcal{ A}} (T) = 0$ or $\mathbb{I} _{\mathcal{B}} (T) = 0$}{\STATE $u \leftarrow T$; }
\ELSE{
\STATE $l \leftarrow T$; }
\ENDIF

\UNTIL{$u <  l + 1$.} \label{alg1:line17}
\end{algorithmic}
\end{algorithm}

\vspace{-0.4cm}
\section{Simulation Results}
In this section, the system performance is evaluated. Consider $M=3$ RNs equally distributed in a cell of radius 750~m. Each RN provides coverage in a radius of 250~m and is located at a distance of 500~m from the BS. We consider $N =$ 5 video files, each of size 500~MB (Bytes), for delivery service to $K=3$ users. The UEs are uniformly and randomly distributed in the cell while the minimum distance between UE and BS/RN is 50~m. Each user requests one file independently. Let $\theta_n$ be the probability of file $n \in \mathcal{N} $ being requested and $\boldsymbol{\theta}=[\theta_1, \ldots, \theta_N]$ be the probability distribution of the requests for the different files. We set $\boldsymbol{\theta}$ = [0.57, 0.20, 0.11, 0.07, 0.05]. Moreover, the path loss model (``Macro + Outdoor Relay, NLOS scenario") in \cite{3GPP:TR36814} is adopted. The small-scale fading coefficients are independent and identically distributed ({i.i.d.}) Rayleigh random variables. Other system parameters are given in Table \ref{tab1}. Before video delivery starts, $\Omega$ = 50~scenarios are randomly generated based on the user preference distribution and the channel model to optimize the initial cache status, cf. (P2a). 

\begin{table}
\small
\centering
\renewcommand{\arraystretch}{1.1}
\protect\caption{Simulation parameters.}
\begin{tabular}{|l|l|}
\hline 
\textbf{Parameters} & \textbf{Settings}\tabularnewline
\hline 
\hline 
System bandwidth & 20 MHz\tabularnewline
\hline 
Subcarriers & $F$ = 64  \tabularnewline
\hline 
Bandwidth of a SC & $W$ = 313 kHz\tabularnewline
\hline  
Duration of time slot & $\Delta$ = 20 ms\tabularnewline
\hline 
Max. transmit power & $P_{S} = $ 46 dBm,  $P_{R} = $ 40 dBm \tabularnewline
\hline 
Noise power density & $N_0$ = $-$172.6 dBm/Hz\tabularnewline
\hline 
Backhaul capacity & $\Gamma_t $ = 1 Gbps, $\forall t$\tabularnewline
\hline 
UE rate requirement & $\nu_{\min}$ = 1 kbps\tabularnewline
\hline 
Initial delay & $\epsilon_{\boldsymbol{\rho}} $ = 0 \tabularnewline
\hline 
\end{tabular}
\vspace{-0.2cm}
\label{tab1}
\end{table}

For comparison, we consider two heuristic caching policies and one suboptimal delivery scheme as baselines:

{\scriptsize$\bullet$} \emph {Baseline 1 (Preference-based Caching):} In this case, the most popular files are cached. Assume that $\boldsymbol{\theta}=[\theta_1, \ldots, \theta_N]$ is known, the cache control decision is made based on
\begin{equation}
\max_{\mathbf{c}^{(m)} \in \mathcal{C}^{(m)}}  \sum\nolimits_{m,n} {\theta_n \cdot c_n^{(m)} \cdot V_n} . 
\label{prefcaching}
\end{equation}

{\scriptsize$\bullet$} \emph{Baseline 2 (Uniform Caching):} In this case, the same amount of data is cached for each file, i.e., $c_n^{(m)} V_n = \frac{1}{N} \times \min\{C_{\max}^{(m)}, \, \sum\nolimits_{n=1}^N V_n\}, \forall m, n$, and the user's preference is not taken into account. For both Baseline 1 and 2, the optimal delivery scheme in (P1a) is adopted.

{\scriptsize$\bullet$} \emph{Baseline 3 (Joint SC Assignment and Power Allocation with Fixed Link Schedule)}: This scheme is basically the same as the one obtained from (P1a) except that a fixed link schedule is assumed, i.e.,  $\mu_{\boldsymbol{\rho},f,2t}^{S} = \mu_{\boldsymbol{\rho},f,2t-1}^{R}=0, \forall t$, holds and hence the benefits of BaR cannot be exploited. Then, joint SC assignment and power allocation are performed for minimizing the delivery time. For Baseline 3, the same initial cache status as for the optimal delivery scheme is adopted.  

\begin{figure}[t]
\centering\includegraphics[scale=0.38]{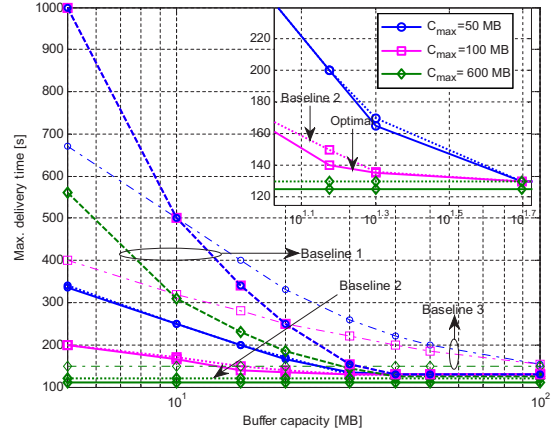}
\caption{Maximum delivery time versus buffer capacity for the proposed scheme (solid line), Baseline~1 (dashed line), Baseline~2 (dotted line), and Baseline~3 (dash-dotted line).}
\label{fig:bfgain}
\vspace{-.4cm}
\end{figure}

\begin{figure}[t]
\centering\includegraphics[scale=0.38]{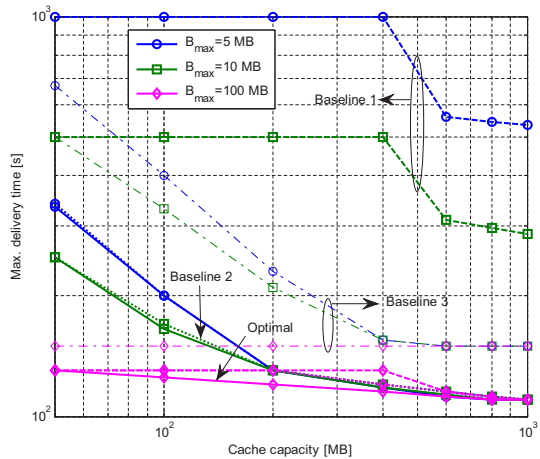}
\caption{Maximum delivery time versus cache capacity for the proposed scheme (solid line), Baseline~1 (dashed line), Baseline~2 (dotted line), and Baseline~3 (dash-dotted line).}
\label{fig:chgain}
\vspace{-.3cm}
\end{figure}

In Figures~\ref{fig:bfgain} and \ref{fig:chgain}, the maximum delivery time of all considered schemes is evaluated for different values of the buffer and cache capacities. For a small cache capacity, we observe from Figure~\ref{fig:bfgain} that the performance of the optimal scheme can be significantly improved by increasing the buffer capacity, which increases the joint scheduling opportunities of (wireless) fetching and delivery for uncached data in the two-hop relaying system. The buffering gains saturate at large buffer capacities when the maximal benefits are achieved. As the cache capacity increases, smaller buffer capacities are sufficient to achieve the maximal buffering gains. This is because the amount of uncached data decreases, which reduces the joint control opportunities for wireless fetching and delivery of uncached data.

The caching gains are further investigated in Figure~\ref{fig:chgain}. For a small buffer capacity, the performance of the optimal scheme improves significantly by increasing the cache capacity. This is expected since the cache facilitates the macroscopic gains of content reuse, traffic diversity, and reduced delivery distance for the delivery of cached data and improves the microscopic diversity gains for uncached data because of joint (wireless and cache) fetching and delivery. Different from the buffering gains, the caching gains do not diminish for large buffer capacities because of macroscopic caching gains which cannot be compensated by buffering gains. Figures~\ref{fig:bfgain} and \ref{fig:chgain} unveil an obvious trade-off between the buffering and the caching gains when the portions of cached and uncached data are changed under different cache and buffer capacities.

From Figures~\ref{fig:bfgain} and~\ref{fig:chgain} we observe that for small cache or buffer capacities, the optimal scheme achieves a gain of about a factor of two in the maximum delivery time compared to Baseline~3 , which confirms the advantages of allocating radio resources through joint scheduling of fetching and delivery. However, when the cache and buffer capacities are sufficiently large, joint SC assignment and power allocation alone can effectively reduce the performance gap. 

\begin{figure}[t]
\centering\includegraphics[scale=0.35]{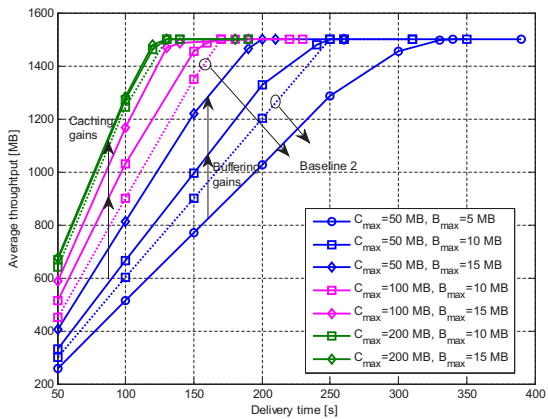}
\caption{Average throughput versus delivery time for the proposed scheme (solid line) and Baseline~2 (dotted line). Results for Baseline 2 are only shown for $B_{\max}$ = 10~MB.}
\label{fig:rt}
\vspace{-.4cm}
\end{figure}

Comparing the optimal scheme with Baseline~1 and~2, our results suggest that preference-based caching is the least efficient in utilizing the cache capacity for overall delivery enhancement, particularly when the cache and buffer capacities are small. The reason is that the uncached files constitute the performance bottleneck. For example, considering (\ref{prefcaching}), the less popular files would not be cached unless $C_{\max} > V_1$, where $V_1$ is the size of the most popular file, i.e., file 1 for the given probability distribution $\boldsymbol{\theta}$. This explains why the performance of Baseline~1 does not improve when $C_{\max}$ increases from 50 MB to 400 MB in Figure~\ref{fig:chgain}. On the other hand, uniform caching combined with the optimal delivery scheme in (P1a), i.e. Baseline 2, performs very close to the optimal scheme. For a detailed analysis of this phenomenon, the average throughput with respect to the delivery time is illustrated in Figure~\ref{fig:rt}. We observe that the average throughput of Baseline~2 increases steadily with the delivery time, because the cache status is independent of the user requests and the delivery process. With both the users' requests and the delivery process considered in the caching decisions, the optimal scheme shows a seemingly faster delivery progress than Baseline~2. However, in an {i.i.d.} channel fading environment, the optimal scheme only improves the delivery completion time slightly compared to Baseline~2.

\vspace{-.2cm}

\section{Conclusion}

In this paper, cross-layer cache and delivery control was investigated for minimizing the video delivery time. A two-stage optimization problem was formulated which turned out to be functional and non-convex. Based on the proposed decomposition and transformation techniques, an efficient algorithm was developed to solve the problem. Simulation results revealed both caching and buffering can effectively improve the delivery performance by exploiting the channel diversity on the fetching and delivery links. Besides, our results unveiled a trade-off between the caching gain and the buffering gain and suggested that uniform caching combined with the proposed optimal delivery scheme can achieve close-to-optimum delivery completion time.

\bibliographystyle{IEEEtran}
\bibliography{IEEEabrv,CellularRelaying,WirelessCaching,WirelessVideo,OptimizationRefs,EnergyHarvesting}

\end{document}